\documentclass[aps,pra,superscriptaddress,twocolumn]{revtex4}
\usepackage{bm}
\usepackage{graphicx,amsmath,latexsym,amssymb}
\usepackage{color}
\usepackage{dcolumn}

\begin{document}
\title{Velocity-dependent dipole forces on an excited atom}
\author{M. Donaire} 
\email{donaire@lkb.upmc.fr, mad37ster@gmail.com}
\affiliation{Laboratoire Kastler Brossel, UPMC-Sorbonnes Universit\'es, CNRS, ENS-PSL Research University,  Coll\`{e}ge de France, 4, place Jussieu, F-75252 Paris, France}
\author{A. Lambrecht}
\affiliation{Laboratoire Kastler Brossel, UPMC-Sorbonnes Universit\'es, CNRS, ENS-PSL Research University,  Coll\`{e}ge de France, 4, place Jussieu, F-75252 Paris, France}

\begin{abstract}
We present a time-dependent calculation of the velocity-dependent  forces which act on an excited atomic dipole in relative motion with respect to ground state atoms of a different kind. Both, its interaction with a single atom and with a dilute atomic plate are evaluated. In either case, the total force consists of a conservative van der Waals  component and a non-conservative R\"ontgen component. On physical grounds, the former corresponds to the velocity-dependent recoil experienced by the excited atom in the processes of absorption and emission of the photons that it exchanges with the ground state atoms on a periodic basis. The latter corresponds to the time-variation of the R\"ontgen momentum, which is also mediated by the periodic exchange of quasi-resonant photons. We find that, at leading order, all these interactions are linear in the velocity.  In the non-retarded regime the van der Waals force dominates, being antiparallel to the velocity. On the contrary, in the retarded regime the velocity-dependent forces oscillate in space, van der Waals and R\"ontgen forces are of the same order in the atom-atom interaction, and the R\"ontgen component dominates in the atom-surface interaction.
\end{abstract}
\maketitle

\section{Introduction}

Frictional forces mediated  by the electromagnetic (EM) vacuum field between neutral atoms and dielectric objects in relative motion has been the subject a number of recent works \cite{Pendry96,Persson,PhilbinUlr,Pendry2010,BuhmannScheel,Barton1,Barton2,BrevikEPJD,IntravaiaDalvit,Intravaiaetal}. As already appreciated by Einstein in his celebrated paper of 1917 \cite{Einstein}, it is the Doppler shift on the photons which interact with a neutral atom in motion with respect to a thermal bath that gives rise to an effective friction which drives the atom to thermal equilibrium. Since then, a number of authors  have discussed the possibility of an analogous effect at zero temperature  \cite{Persson,PhilbinUlr,Pendry2010,BuhmannScheel,Barton1,Barton2,BrevikEPJD,IntravaiaDalvit,Intravaiaetal,Echenique,Levitov,Barton96,Brevik92,Pendry96}. In this case, it is the exchange of vacuum photons between the fluctuating currents of neutral objects moving at relative velocities that gives rise to a net transfer of momentum between them in the direction parallel to the velocity. In this respect, the explanatory work by Pendry \cite{Pendry96} has motivated the investigation of the so-called \emph{quantum friction} in different scenarios. Several approaches and approximations have been considered since then by a number of authors, yielding apparent contradictory results with regard to the dependence of the friction force on the velocity. Nonetheless, recent works agree in the cubic scaling of the friction force with velocity on a ground state atom at zero temperature  in motion with respect to a metallic slab \cite{BrevikEPJD,IntravaiaDalvit,Intravaiaetal}. For the case of the interaction between two macroscopic objects the approaches are semiclassical and based on linear response theory and the fluctuation dissipation theorem \cite{Persson,PhilbinUlr}. Concerning the interaction between an atom and a dielectric object, the authors of Ref.\cite{Echenique,BuhmannScheel,IntravaiaDalvit} have treated the atom quantum-mechanically, while the electric response of the system macroscopic object-EM field  has been considered semiclassical. On the contrary, the authors of Refs.\cite{Barton96,Barton2,Intravaiaetal} have applied time-dependent perturbation theory within a Hamiltonian approach in which the interaction between the macroscopic objects and the EM field is effective and semiclassical. Lastly, the interaction between two harmonic oscillators (intended to describe two atoms) has been considered by the authors of Ref.\cite{Barton1,BrevikEPJD,Brevik92} within a Hamiltonian approach in which the EM interaction between them reduces to the electrostatic (i.e., non-quantum) one.  Finally, all the aforementioned works, except for that by Scheel and Buhmann \cite{BuhmannScheel}, deal with  bodies in their (internal) ground states, and they all appeal to some effective mechanism of dissipation to explain quantum friction.
\begin{figure}[h]
\includegraphics[height=5.0cm,width=8.9cm,clip]{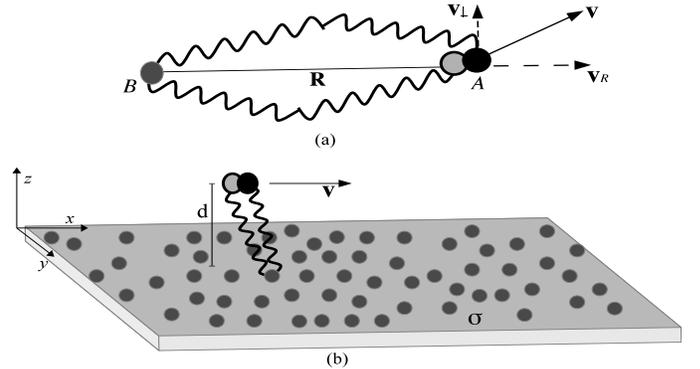}
\caption{(a) Sketch of the interaction of atom $A$ with an atom $B$. Atom $A$ moves at constant velocity $\mathbf{v}$ while atom $B$ remains at rest at a distance $R$ which varies in time. (b) Sketch of the interaction of an atom $A$ with a thin plate made of a random distribution of  atoms of kind $B$, with  numerical surface density $\sigma$. The atom $A$ flies with velocity $\mathbf{v}$ parallel to the plate at a height $d$. The wavy lines represent the pairs of photons exchanged by the atoms. Emission and absorption processes take place at different locations of atom $A$.}\label{figure1}
\end{figure}

In this article we study the velocity-dependent EM force between two neutral bodies, one of which is initially prepared in an excited state. In the first place we study  the interaction between two  atoms of different kinds, $A$ and $B$, in relative motion, one of which, say atom $A$, is initially prepared in an excited state at zero temperature--see Fig.\ref{figure1}(a). This is probably the simplest realistic scenario where a Hamiltonian treatment can be carried out all the way to the end. Important is the fact that dissipation is purely radiative here, which allows us to trace it along the calculation. In order to simplify the calculation we adopt the \emph{quasiresonant approximation} outlined in Ref.\cite{PRLvdW}. That is, we assume two-level atoms for which the detuning between their resonant frequencies, $\Delta_{AB}\equiv\omega_{A}-\omega_{B}$, is such that $\Gamma_{A,B}<\Delta_{AB}\ll\omega_{A,B}$, where $\Gamma_{A,B}$ are the linewidths of the corresponding transitions of atoms $A$ and $B$ respectively. This condition is easily met in pairs of alkali atoms as well as, in an approximate manner, in pairs of circular Rydberg atoms \cite{Raimond}. It allows us to truncate our results at leading order in $\Delta_{A,B}/\omega_{AB}$. In addition, we assume that the observation time $T$ satisfies $2\pi/|\Delta_{AB}|<T\ll2\pi/\Gamma_{A,B}$,  such  that atom $A$ remains excited until the obervation time.  We will extend our approach to the calculation of the velocity-dependent force on an excited atom of kind $A$ flying at constant velocity parallel to a thin plate made of a random distribution of atoms of kind $B$ in their ground state --see Fig.\ref{figure1}(b).  

The article is organized as follows. In Sec.\ref{Fundamental} we explain the fundamentals of our approach. In Sec.\ref{vdW} we compute the velocity-dependent van der Waals force. The R\"ontgen force is evaluated in Sec.\ref{Rontgen}. Conclusions are summarized in Sec.\ref{Conclusion}.

\section{Fundamentals of the calculation}\label{Fundamental}

We aim at computing the velocity-dependent force on the atom $A$ as a result of its EM interaction and its relative motion with respect to atom $B$, in the electric-dipole and quasiresonant approximations. Our approach is based on that of Ref.\cite{PRLvdW}, which is appropriately extended to account for dynamical effects. Basically, the extention consists of, in the first place,  promoting the center of mass (CM) degrees of freedom (d.o.f.) of the atoms to quantum operators, and secondarily, adding the contribution of the R\"ontgen momentum both to the force operator and to the interaction potential. Following Ref.\cite{Baxteretal},  
in the electric-dipole approximation the canonical conjugate momenta of the atomic CM position vectors $\mathbf{R}_{A,B}$ are $\mathbf{P}_{A,B}=\mathbf{Q}_{A,B}-\mathbf{d}_{A,B}\cdot\mathbf{B}(\mathbf{R}_{A,B})$, $\mathbf{Q}_{A,B}$ being the kinetic momenta of the CM of each atom. The interaction potential between the free EM field and the atomic dipoles is $W=W^{E}_{A}+W^{E}_{B}+W^{R}_{A}+W^{R}_{B}$, where 
\begin{eqnarray}
W^{E}_{A,B}&=&-\mathbf{d}_{A,B}\cdot\mathbf{E}(\mathbf{R}_{A,B}),\nonumber\\
W^{R}_{A,B}&=&\Bigl[\mathbf{P}_{A,B}\cdot[\mathbf{d}_{A,B}\times\mathbf{B}(\mathbf{R}_{A,B})]\nonumber\\
&+&[\mathbf{d}_{A,B}\times\mathbf{B}(\mathbf{R}_{A,B})]\cdot\mathbf{P}_{A,B}\Bigr]/2m_{A,B}\nonumber
\end{eqnarray}
are the electric field and R\"ontgen coupling potentials respectively which act upon the free Hamiltonians. These are, the free Hamiltonian of the atomic internal d.o.f., $H_{A}+H_{B}=\hbar\omega_{A}|A_{+}\rangle\langle A_{+}|+\hbar\omega_{B}|B_{+}\rangle\langle B_{+}|$, the kinetic energy of the atomic CMs, $\mathcal{K}_{A,B}=c\sqrt{Q_{A,B}^{2}+c^{2}m^{2}_{A,B}}$, and the Hamiltonian of free photons,  $H_{EM}=\sum_{\mathbf{k},\mathbf{\epsilon}}\hbar\omega(a_{\mathbf{k},\mathbf{\epsilon}}a^{\dagger}_{\mathbf{k},\mathbf{\epsilon}}+1/2)$.  In the above expresssions $\mathbf{d}_{A,B}$ denote the electric dipole moment operators of the atoms $A$ and $B$ respectively, $m_{A,B}$ are their respective atomic masses, and $\mathbf{E}$, $\mathbf{B}$ are the electric and magnetic field operators  which can be expressed as series of annihilation and creation field operators of photons of momentum $\hbar\mathbf{k}$, respectively,
\begin{eqnarray}\label{AQ}
\mathbf{E}(\mathbf{R}_{A,B})&=&\sum_{\mathbf{k}}\mathbf{E}^{(-)}_{\mathbf{k}}(\mathbf{R}_{A,B})+\mathbf{E}^{(+)}_{\mathbf{k}}(\mathbf{R}_{A,B})\nonumber\\
&=&i\sum_{\mathbf{k},\mathbf{\epsilon}}\sqrt{\frac{\hbar ck}{2\mathcal{V}\epsilon_{0}}}
[\mathbf{\epsilon}a_{\mathbf{k},\epsilon}e^{i\mathbf{k}\cdot\mathbf{R}_{A,B}}-\mathbf{\epsilon}^{*}a^{\dagger}_{\mathbf{k},\epsilon}e^{-i\mathbf{k}\cdot\mathbf{R}_{A,B}}],\nonumber\\
\mathbf{B}(\mathbf{R}_{A,B})&=&\sum_{\mathbf{k}}\mathbf{B}^{(-)}_{\mathbf{k}}(\mathbf{R}_{A,B})+\mathbf{B}^{(+)}_{\mathbf{k}}(\mathbf{R}_{A,B})\nonumber\\
&=&i\sum_{\mathbf{k},\mathbf{\epsilon}}\sqrt{\frac{\hbar}{2ck\mathcal{V}\epsilon_{0}}}\mathbf{k}\times
[\mathbf{\epsilon}a_{\mathbf{k},\epsilon}e^{i\mathbf{k}\cdot\mathbf{R}_{A,B}}-\mathbf{\epsilon}^{*}a^{\dagger}_{\mathbf{k},\epsilon}e^{-i\mathbf{k}\cdot\mathbf{R}_{A,B}}],\nonumber
\end{eqnarray}
where $\mathcal{V}$ is a quantisation volume, $\omega=ck$ is the photon frequency, and the operators $a^{\dagger}_{\mathbf{k},\mathbf{\epsilon}}$ and $a_{\mathbf{k},\mathbf{\epsilon}}$ are the creation and annihilation operators of photons of momentum $\hbar\mathbf{k}$ and polarization $\mathbf{\epsilon}$ respectively.

 As in Ref.\cite{PRLvdW}, we use the wave function formalism in the Schr\"odinger picture \cite{Shankar} to compute the velocity dependent-force on an excited atom $A$ which moves at certain velocity $\mathbf{v}$ with respect to (w.r.t.) an atom $B$ in its ground state. That is, denoting by $\mathbb{U}(T)$ the time-evolution operator at the time of observation $T$, $\mathbb{U}(T)=\mathcal{T}\exp{}\Bigl\{-i\hbar^{-1}\int_{0}^{T}\textrm{d}t\:[H_{A}+H_{B}+H_{EM}+\mathcal{K}_{A}+\mathcal{K}_{B}+W^{E}_{A}+W^{E}_{B}+W^{R}_{A}+W^{R}_{B}]\Bigr\}$, and by $|\Psi(0)\rangle$ the wave function of the two-atom system right at the moment at which atom $A$ gets excited,  the total force acting upon atom $A$ at $T>0$ reads, at $\mathcal{O}[(W_{A}^{E}+W_{A}^{R})^{2}]$,
\begin{align}
\langle\mathbf{F}_{A}(T)\rangle&=\partial_{T}\langle\mathbf{Q}_{A}(T)\rangle=-i\hbar\partial_{T}\langle\Psi(0)|\mathbb{U}^{\dagger}(T)\mathbf{\nabla}_{\mathbf{R}_{A}}\mathbb{U}(T)|\Psi(0)\rangle\nonumber\\
&+\partial_{T}\langle\Psi(0)|\mathbb{U}^{\dagger}(T)\mathbf{d}_{A}\times\mathbf{B}(\mathbf{R}_{A})\mathbb{U}(T)|\Psi(0)\rangle\nonumber\\
&\simeq-\mathbf{\nabla}_{\mathbf{R}_{T}}\langle[W^{E}_{A}+W^{R}_{A}](T)\rangle/2+\partial_{T}\langle[\mathbf{d}_{A}\times\mathbf{B}(\mathbf{R}_{A})](T)\rangle\nonumber\\
&\simeq\frac{1}{2}\mathbf{\nabla}_{\mathbf{R}_{T}}\langle\mathbf{d}_{A}\cdot[\mathbf{E}(\mathbf{R}_{A})+m_{A}^{-1}\langle\mathbf{Q}_{A}\rangle\times\mathbf{B}(\mathbf{R}_{A})](T)\rangle\nonumber\\
&+\partial_{T}\langle[\mathbf{d}_{A}\times\mathbf{B}(\mathbf{R}_{A})](T)\rangle.\label{Force}
\end{align}
In this equation a term quadratic in $\mathbf{B}(\mathbf{R}_{A})$ has been discarded in the last line as it yields a negligible contribution under quasiresonant conditions and we assume that quantum fluctuations over the classical trajectories of the atoms are negligible, both in the positions and in the kinetic momenta of their CMs. That is, denoting by $\mathbb{U}_{\mathcal{K}}(T)=\exp{}\{-i\hbar^{-1}[\mathcal{K}_{A}+\mathcal{K}_{B}]\}$ the free-evolution operator of the CM d.o.f., the (approximately) constant values of the atomic kinetic momenta 
read $\langle \mathbf{Q}_{A,B}\rangle\equiv\langle\Psi(0)|\mathbb{U}^{\dagger}_{\mathcal{K}}(t)\mathbf{Q}_{A,B}\mathbb{U}_{\mathcal{K}}(t)|\Psi(0)\rangle\:\forall\:t$. Next, neglecting quantum fluctuations over the expectation values of the CM position and momentum vectors, we can write $\langle\Psi(0)|\mathbb{U}^{\dagger}_{\mathcal{K}}(t)\mathcal{F}(\mathbf{R}_{A,B},\mathbf{Q}_{A,B})\mathbb{U}_{\mathcal{K}}(t)|\Psi(0)\rangle\simeq\mathcal{F}(\langle\mathbf{R}_{A,B}(t)\rangle,\langle\mathbf{Q}_{A,B}\rangle)$ for  any functional  $\mathcal{F}$ of the operators $\mathbf{R}_{A,B}$ and $\mathbf{Q}_{A,B}$. In particular, the functional gradient $\mathbf{\nabla}_{\mathbf{R}_{A}}$ is replaced with $\mathbf{\nabla}_{\mathbf{R}_{T}}$ in Eq.(\ref{Force}), with $\mathbf{R}_{T}\equiv\langle[\mathbf{R}_{A}-\mathbf{R}_{B}](T)\rangle$, and the constant velocity of atom $A$ relates to $\langle\mathbf{Q}_{A}\rangle$ through the classical relativistic expression $\mathbf{v}=\frac{\partial}{\partial t}\langle\Psi(0)|\mathbb{U}_{\mathcal{K}}^{\dagger}(t)\mathbf{R}_{A}\mathbb{U}_{\mathcal{K}}(t)|\Psi(0)\rangle\simeq\frac{\langle\mathbf{Q}_{A}\rangle/m_{A}}{\sqrt{1+\langle Q_{A}\rangle^{2}/(m_{A}c)^{2}}}$.

Similar to the calculation carried out in Ref.\cite{PRLvdW} for the van der Waals (vdW) interaction, we will apply standard time-dependent perturbation theory at order 4 in $W$ to compute the force of Eq.(\ref{Force}) under quasiresonant conditions. The only difference here is the additional coupling to the R\"ontgen momentum. We will address first the velocity-dependent force in the absence of the R\"ontgen momentum, which we will refer to as velocity-dependent vdW force. Later, we will compute the terms of the force which depend on the R\"ontgen momentum, which we will refer to as R\"ontgen force for brevity.

\section{Velocity-dependent vdW force}\label{vdW}

\begin{figure}[h]
\includegraphics[height=4.0cm,width=8.4cm,clip]{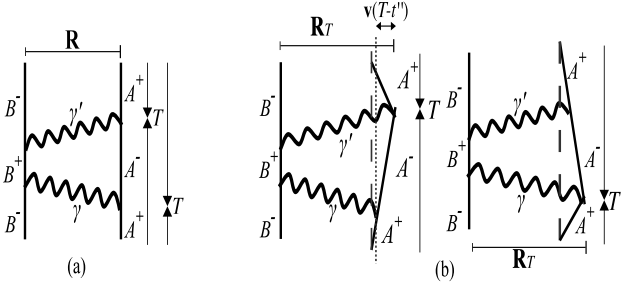}
\caption{(a) Diagrammatic representation of  Eq.(\ref{laeq}) for $\langle W^{E}_{A}(T)\rangle$. Time runs along the vertical as indicated by the arrows. $\mathbf{R}$ is intended, generally, as a quantum operator. Only for both atoms at rest, $\mathbf{R}$ can be treated in this picture as a classical variable. (b) Diagrammatic representation of  Eq.(\ref{Wv}) for $\langle W^{E}_{A}(T)\rangle^{\mathbf{v}}$ up to order $(v/c)^{3}$.  The interatomic displacement is here a classical variable which depends on time. $\mathbf{v}(T-t'')$ is the displacement of atom $A$ during the lag between the emission of the first photon and the absorbtion of the second photon.}\label{figure2}
\end{figure}

It was found in Ref.\cite{PRLvdW}, using standard time-dependent perturbation theory at order 4 in the potential $W^{E}_{A}+W^{E}_{B}$ that the  energy of interaction between an excited atom $A$ and a ground state atom $B$ at time $T$ after the excitation of atom $A$ is given by 
\begin{widetext}
\begin{align}\label{laeq}
\langle W^{E}_{A}(T)\rangle&\simeq\frac{1}{\hbar^{3}}\int_{-\infty}^{\infty}\frac{\mathcal{V}k^{2}\textrm{d}k}{(2\pi)^{3}}\int_{-\infty}^{\infty}\frac{\mathcal{V}k^{'2}\textrm{d}k'}{(2\pi)^{3}}\int_{0}^{4\pi}\textrm{d}\Omega\int_{0}^{4\pi}\textrm{d}\Omega'\Bigl[i\langle\Psi(0)|\mathbb{U}_{0}(-T)|\Psi(0)\rangle\Theta(T-2R/c)\nonumber\\
&\times\int_{0}^{T}\textrm{d}t\int_{0}^{t}\textrm{d}t'\int_{0}^{t'}\textrm{d}t''
\langle\Psi(0)|\mathbf{d}_{A}\cdot\mathbf{E}_{\mathbf{k}'}^{(-)}(\mathbf{R}_{A})\mathbb{U}_{0}(T-t)\mathbf{d}_{B}\cdot\mathbf{E}_{\mathbf{k}'}^{(+)}(\mathbf{R}_{B})\mathbb{U}_{0}(t-t')\mathbf{d}_{B}\cdot\mathbf{E}_{\mathbf{k}}^{(-)}(\mathbf{R}_{B})\nonumber\\&\times\mathbb{U}_{0}(t'-t'')\mathbf{d}_{A}\cdot\mathbf{E}_{\mathbf{k}}^{(+)}(\mathbf{R}_{A})\mathbb{U}_{0}(t'')|\Psi(0)\rangle\Bigr]+[k\leftrightarrow k']^{\dagger},
\end{align}
\end{widetext} 
where $\mathbf{R}$ the interatomic displacement vector, $\mathbf{R}=\mathbf{R}_{A}-\mathbf{R}_{B}$ and $\mathbb{U}_{0}(t)$ is the  unperturbed time-evolution propagator in the absence of $W$. Eq.(\ref{laeq}) corresponds almost completely to the diagram of Fig.\ref{figure2}(a) alone, where the photonic loop is made of two doubly-resonant photons. In addition, the diagrams $(b)-(f)$ of figure 1 in Ref.\cite{PRLvdW} provide the Heaviside function $\Theta(T-2R/c)$ which guarantees causality.

For both atoms at rest (which we denote by the superscript $\mathbf{0}$) their position vectors $\mathbf{R}_{A}$ and $\mathbf{R}_{B}$ can be considered classical and constant vectors, and  $\mathbb{U}_{0}(t)$ reads $\mathbb{U}^{\mathbf{0}}_{0}(t)=\exp{[-i\hbar^{-1}(H_{A}+H_{B}+H_{EM})t]}$. 
Finally, $|\Psi(0)\rangle$ at zero velocity is made of the tensor product of the EM vacuum state ($|0_{\gamma}\rangle$) and the internal states of the atoms, the excited one for atom $A$  and the fundamental one for atom $B$,  $|\Psi^{\mathbf{0}}(0)\rangle=|A_{+}\rangle\otimes|B_{-}\rangle\otimes|0_{\gamma}\rangle$. Substituting next these formulas in Eq.(\ref{laeq}), taking the limit to the continuum by replacing the sums over momenta by integrals, summing over polarizations and integrating in orientations we arrive at \cite{PRLvdW}
\begin{align}
\langle W^{E}_{A}(T)\rangle&^{\mathbf{0}}\simeq\frac{4\alpha_{f} c^{3}}{\pi\epsilon_{0}e^{2}}\mu^{A}_{i}\mu^{B}_{j}\mu^{B}_{p}\mu^{A}_{q}\int_{-\infty}^{+\infty}\textrm{d}k\:k^{2}\textrm{Im}\{G^{ij}(k'R)\}\nonumber\\
&\times\int_{-\infty}^{+\infty}\textrm{d}k'\:k'^{2}\textrm{Im}\{G^{pq}(kR)\}\Theta(T-2R/c)\label{W0}\\
&\times\int_{0}^{T}\textrm{d}t\int_{0}^{t}\textrm{d}t'\int_{0}^{t'}\textrm{d}t''\Bigl[\bigl(i\:e^{i\omega_{A}T}e^{-i(T-t)\omega'}\nonumber\\
&\times e^{-i(t-t')\omega_{B}}e^{-i(t'-t'')\omega}e^{-it''\omega_{A}}\bigr)+(\omega\leftrightarrow\omega')^{\dagger}\Bigr],\nonumber
\end{align}
where $\alpha_{f}$ is the fine structure constant, $\mu^{A}=\langle A_{-}|\mathbf{d}_{A}|A_{+}\rangle$, $\mu^{B}=\langle B_{-}|\mathbf{d}_{B}|B_{+}\rangle$,  and the tensor $\mathbb{G}(kR)$ is the dyadic Green's function of the electric field induced by an electric dipole of frequency $ck$ in free space,
\begin{equation}
\mathbb{G}(kR)=\frac{k\:e^{ikR}}{4\pi}[\alpha/kR+i\beta/(kR)^{2}-\beta/(kR)^{3}],
\end{equation}
where the tensors $\alpha$ and $\beta$ read $\alpha=\mathbb{I}-\mathbf{R}\mathbf{R}/R^{2}$,  $\beta=\mathbb{I}-3\mathbf{R}\mathbf{R}/R^{2}$.

Let us consider next that the atoms move away from each other at relative velocity $\mathbf{v}$. For simplicity, we consider atom $B$ at rest in our reference frame. We will denote the associated observables by the superscript $\mathbf{v}$. At leading order, the dominant contribution to the interaction still comes from Eq.(\ref{laeq}), but now allowing for the constant motion of the CM of each atom. In order to account for this motion, the position vectors of the CMs, $\mathbf{R}_{A,B}$,  and their kinetic momenta,   $\mathbf{Q}_{A,B}$, are promoted to quantum operators; the initial state is now the tensor product of the EM vacuum, the initial internal atomic states and the states of the CMs of each atom;  and the time-evolution propagator contains the kinetic energy of the CMs defined earlier, $\mathbb{U}_{0}(t)=\mathbb{U}^{\mathbf{0}}_{0}(t)\otimes\mathbb{U}_{\mathcal{K}}(t)$. When the CM d.o.f. are considered quantum operators in Eq.(\ref{laeq}), the resultant equation reads
\begin{align}
&\langle W^{E}_{A}(T)\rangle^{\mathbf{v}}\simeq\nonumber\\
&\int\textrm{d}t\textrm{d}t'\textrm{d}t''\textrm{d}k\textrm{d}k'[\langle\Psi(0)|\mathcal{C}(A_{\pm},B_{\pm},\gamma;k,k',t,t',t'',T)\nonumber\\&\times\mathbb{U}_{\mathcal{K}}(-T)e^{i\mathbf{k}'\cdot\mathbf{R}_{A}}\mathbb{U}_{\mathcal{K}}(T-t)e^{-i\mathbf{k}'\cdot\mathbf{R}_{B}}\mathbb{U}_{\mathcal{K}}(t-t')e^{i\mathbf{k}\cdot\mathbf{R}_{B}}\nonumber\\&\times\mathbb{U}_{\mathcal{K}}(t'-t'')e^{-i\mathbf{k}\cdot\mathbf{R}_{A}}\mathbb{U}_{\mathcal{K}}(t'')|\Psi(0)\rangle]+[k\leftrightarrow k']^{\dagger},\label{laeq1}
\end{align}
where we have isolated explicitly the factors which depend on the CM operators alone, with $\mathcal{C}$ being a function of the photonic and internal atomic d.o.f. as well as of the time and frequency variables. 
In the absence of the R\"ontgen momentum, the following relation 
holds in application of the canonical commutaion relations, $[R^{i}_{A,B},Q^{j}_{A,B}]\simeq i\hbar\delta^{ij}$, \footnote{The inclusion of the R\"ontgen momentum would yield contributions of higher order in $W$, hence being negligible.}
\begin{align}\label{exp}
&e^{-i\mathbf{k}\cdot\mathbf{R}_{A,B}}e^{-i\mathcal{K}(\mathbf{Q}_{A,B})t/\hbar}e^{i\mathbf{k}\cdot\mathbf{R}_{A,B}}=e^{-i\mathcal{K}(\mathbf{Q}_{A,B}+\hbar\mathbf{k})t/\hbar}\nonumber\\
&\simeq e^{-i\mathcal{K}(\mathbf{Q}_{A,B})t/\hbar}\nonumber\\
&\times\exp{\left[\frac{-i\mathbf{k}\cdot\mathbf{Q}_{A,B}}{m_{A,B}}t[1-\frac{Q_{A,B}^{2}/2}{(m_{A,B}c)^{2}}+\frac{3Q_{A,B}^{4}/8}{(m_{A,B}c)^{4}}]\right]}.
\end{align}
In the last line we have expanded $\mathcal{K}$ up to terms of order 5 in $Q_{A,B}/m_{A,B}c$. Applying the above relation to Eq.(\ref{laeq1}) we end up with the expression,
\begin{align}
&\langle W^{E}_{A}(T)\rangle^{\mathbf{v}}\simeq\int\textrm{d}t\textrm{d}t'\textrm{d}t''\textrm{d}k\textrm{d}k'\nonumber\\
&\times\Bigl\{\langle\Psi(0)|\mathbb{U}_{\mathcal{K}}(-t')\mathcal{C}(A_{\pm},B_{\pm},\gamma;k,k',t,t',t'',T)\nonumber\\&\times\exp{\left[i\mathbf{k}'\cdot\frac{\mathbf{Q}_{A}}{m_{A}}(T-t')[1-\frac{Q_{A}^{2}}{2(m_{A}c)^{2}}+\frac{3Q_{A}^{4}}{8(m_{A}c)^{4}}]\right]}\nonumber\\&\times\exp{\left[-i\mathbf{k}'\cdot\frac{\mathbf{Q}_{B}}{m_{B}}(t-t')[1-\frac{Q_{B}^{2}}{2(m_{B}c)^{2}}+\frac{3Q_{B}^{4}}{8(m_{B}c)^{4}}]\right]}\nonumber\\&\times e^{i\mathbf{k}'\cdot(\mathbf{R}_{A}-\mathbf{R}_{B})} e^{-i\mathbf{k}\cdot(\mathbf{R}_{A}-\mathbf{R}_{B})}\nonumber\\&
\times\exp{\left[i\mathbf{k}\cdot\frac{\mathbf{Q}_{A}}{m_{A}}(t'-t'')[1-\frac{Q_{A}^{2}}{2(m_{A}c)^{2}}+\frac{3Q_{A}^{4}}{8(m_{A}c)^{4}}]\right]}\nonumber\\&
\times\mathbb{U}_{\mathcal{K}}(t')|\Psi(0)\rangle\Bigr\}+\{k\leftrightarrow k'\}^{\dagger},\label{laeq2}
\end{align}
where terms of the order of $\hbar k/Q_{A,B}$ times less have been ignored in the exponentials \footnote{This approximation relies on the assumption that the active electrons within the atoms are non-relativistic, which implies an implicit  wavelength cut-off of the order of the  electronic Compton wavelenght, which translates into the assumption $m_{e}c/m_{A}v\ll1$.}. 
We observe that the factors which depend on the CM d.o.f. are evaluated on the state of the CM at time $t'$, $\mathbb{U}_{\mathcal{K}}(t')|\Psi(0)\rangle$. Neglecting quantum fluctuations in the CM d.o.f. and making use of the approximation
\begin{equation}\label{eq7}
\mathbf{v}\simeq\frac{\langle\mathbf{Q}_{A}\rangle}{m_{A}}\left[1-\frac{\langle Q_{A}\rangle^{2}}{2(m_{A}c)^{2}}+\frac{\langle Q_{A}\rangle^{4}}{8(m_{A}c)^{4}}\right],
\end{equation}
we find
\begin{align}
&\langle W^{E}_{A}(T)\rangle^{\mathbf{v}}\simeq\int\textrm{d}t\textrm{d}t'\textrm{d}t''\textrm{d}k\textrm{d}k'\nonumber\\
&\times[\langle\Psi(0)|\mathcal{C}(A_{\pm},B_{\pm},\gamma;k,k',t,t',t'',T)|\Psi(0)\rangle\nonumber\\&\times
e^{-i\mathbf{k}\cdot[\langle\mathbf{R}_{A}(T)\rangle-\mathbf{v}(T-t'')-\langle\mathbf{R}_{B}\rangle]}
e^{i\mathbf{k}'\cdot[\langle\mathbf{R}_{A}(T)\rangle-\langle\mathbf{R}_{B}\rangle]} \label{Doppy}\\
&\times e^{-i\mathbf{k}\cdot\mathbf{v}(v/c)^{4}(t''-t')/4}e^{i\mathbf{k}'\cdot\mathbf{v}(v/c)^{4}(T-t')/4}]+[k\leftrightarrow k']^{\dagger}.\nonumber
\end{align}
This result shows that, up to terms of order 3 in $v/c$, the net effect of the application of Eqs. (\ref{exp}) and (\ref{eq7}) is the substitution of the operators $\mathbf{R}_{A,B}$ in Eq.(\ref{laeq}) by their corresponding classical vectors which evolve in time as $\langle\mathbf{R}_{B}(t)\rangle=\langle\mathbf{R}_{B}\rangle$ constant, $\langle\mathbf{R}_{A}(t)\rangle=\langle\mathbf{R}_{A}(T)\rangle-\mathbf{v}(T-t)$, with $0\leq t\leq T$. The absence of Doppler shifts of order $(v/c)^{3}$ is remarkable.  


The resultant diagrammatic representation of $\langle W^{E}_{A}(T)\rangle^{\mathbf{v}}$ up to $\mathcal{O}[(v/c)^{3}]$  is shown in Fig.\ref{figure2}(b). Once integrated in orientations, the interaction energy of the moving atom $A$ reads,
\begin{align}
\langle W^{E}_{A}(T)&\rangle^{\mathbf{v}}\simeq\frac{4\alpha_{f}c^{3}}{\pi\epsilon_{0}e^{2}}\mu^{A}_{i}\mu^{B}_{j}\mu^{B}_{p}\mu^{A}_{q}\int_{-\infty}^{+\infty}\textrm{d}k'\int_{-\infty}^{+\infty}\textrm{d}k\nonumber\\
&\times\Bigl\{\int_{0}^{T}\textrm{d}t\int_{0}^{t}\textrm{d}t'\int_{0}^{t'}\textrm{d}t''\Bigl[k'^{2}\textrm{Im}\{G^{ij}(k'R_{T})\}\nonumber\\
&\times k^{2}\textrm{Im}\{G^{pq}[k|\mathbf{R}_{T}-\mathbf{v}(T-t'')|]\}\label{Wv}\\
&\times\Theta\left[T(1+v_{R}/c)-2R_{T}/c\right]\bigl(i\:e^{i\omega_{A}T}e^{-i(T-t)\omega'}\nonumber\\
&\times e^{-i(t-t')\omega_{B}}e^{-i(t'-t'')\omega}e^{-it''\omega_{A}}\bigr)\Bigr]+[\omega\leftrightarrow\omega']^{\dagger}\Bigr\},\nonumber
\end{align}
where $\mathbf{R}_{T}$ was defined earlier as the interatomic displacement vector at the time of observation. In Eq.(\ref{Wv}) the argument of the first Green function is shifted w.r.t. the one in Eq.(\ref{W0}) by an amount proportional to the displacement of atom $A$ during the lag between the emission of the first photon and the absorbtion of the second photon by atom $A$, $\mathbf{v}(T-t'')$. Correspondingly,  the argument of the Heaviside function is shifted by an amount $v_{R}T/c$, where $v_{R}$ is the component of $\mathbf{v}$ parallel to $\mathbf{R}_{T}$. The shift in the argument of the Green function gives rise to two kinds of terms at $\mathcal{O}(v_{R}/c)$,
\begin{widetext}
\begin{align}
&\mathbb{G}[k|\mathbf{R}_{T}-\mathbf{v}(T-t'')|]\simeq\frac{k\:e^{ik(R_{T}-v_{R}T+v_{R}t'')}}{4\pi}\Bigl[\frac{\alpha_{T}}{k[R_{T}-v_{R}(T-t'')]}+\frac{i\beta_{T}}{[k(R_{T}-v_{R}(T-t''))]^{2}}-\frac{\beta_{T}}{[k(R_{T}-v_{R}(T-t''))]^{3}}\Bigr]\nonumber\\
&\simeq\frac{k\:e^{ikR_{T}}}{4\pi}\Bigl[e^{-ikv_{R}(T-t'')}[\frac{\alpha_{T}}{kR_{T}}+\frac{i\beta_{T}}{(kR_{T})^{2}}-\frac{\beta_{T}}{(kR_{T})^{3}}]\:+\:\frac{v_{R}(T-t'')}{R_{T}}[\frac{\alpha_{T}}{kR_{T}}+\frac{2i\beta_{T}}{(kR_{T})^{2}}-\frac{3\beta_{T}}{(kR_{T})^{3}}]\Bigr],\label{GDoppLag}
\end{align}
\end{widetext}
where the tensors read $\alpha_{T}=\mathbb{I}-\mathbf{R}_{T}\mathbf{R}_{T}/R_{T}^{2}$, $\beta_{T}=\mathbb{I}-3\mathbf{R}_{T}\mathbf{R}_{T}/R_{T}^{2}$.
The  terms within the first small square brackets of the second line of Eq.(\ref{GDoppLag}) can be interpreted as  Doppler shifts on the frequency of the photon emitted from atom $A$, whereas the terms within the second small square brackets are multiplied by the lag between the photons emitted and  absorbed by atom $A$. This distinction is however merely formal, since in both cases the origin of these terms lies in the Doppler shifts of the complex exponentials of Eq.(\ref{Doppy}), restricted to terms linear in $v_{R}/c$. 

The subtitution of Eq.(\ref{GDoppLag}) in Eq.(\ref{Wv}) yields three different contributions to $\langle W^{E}_{A}(T)\rangle^{\mathbf{v}}$. We will denote the interaction associated to the first term of Eq.(\ref{GDoppLag}) by $\langle W^{E}_{A}(T)\rangle^{\mathbf{v}}_{Dop}$, and  the interaction associated to the second term of Eq.(\ref{GDoppLag}) by $\langle W_{A}(T)\rangle^{\mathbf{v}}_{Lag}$. Lastly, the Heaviside function contributes at leading order with a term $\langle W^{E}_{A}\rangle^{\mathbf{v}}_{\Theta}=\frac{-v_{R}\partial}{2\partial v_{R}}\langle W^{E}_{A}[2R_{T}(1-v_{R}/c)/c]\rangle^{\mathbf{0}}|_{v_{R}=0}$. In summary, we obtain
\begin{align}
\langle W^{E}_{A}(T)\rangle^{\mathbf{v}}_{\Theta}&=\frac{2\mathcal{U}_{ijpq}\Delta_{AB}v_{R}}{R_{T}^{5}c^{2}}\bigl[\beta_{T}^{ij}\beta_{T}^{pq}-k_{B}^{2}R_{T}^{2}(\beta_{T}^{ij}\beta_{T}^{pq}+2\alpha_{T}^{ij}\beta_{T}^{pq})\nonumber\\
&+k_{B}^{4}R_{T}^{4}\alpha_{T}^{ij}\alpha_{T}^{pq}\bigr]\sin{(2k_{A}R_{T})}-\frac{4\mathcal{U}_{ijpq}\Delta_{AB}v_{R}}{R_{T}^{4}c^{2}}k_{B}\nonumber\\&\times\left[\beta_{T}^{ij}\beta_{T}^{pq}-k_{B}^{2}R_{T}^{2}\alpha_{T}^{ij}\beta_{T}^{pq}\right]\cos{(2k_{A}R_{T})},
\end{align}
which is of the order of $v_{R}\Delta_{AB}R_{T}/c^{2}$ times the interaction $\langle W^{E}_{A}(T)\rangle^{\mathbf{0}}$ calculated in Ref.\cite{PRLvdW}, with $\mathcal{U}_{ijpq}=\mu^{A}_{i}\mu^{A}_{q}\mu^{B}_{j}\mu^{B}_{p}/[(4\pi\epsilon_{0})^{2}\hbar\Delta_{AB}]$. As for the Doppler term we find
\begin{widetext}
\begin{align}\label{Dop}
\langle W_{A}(T)\rangle^{\mathbf{v}}_{Dop}&=\frac{2\tilde{\mathcal{U}}_{ijpq}}{R_{T}^{6}}[\beta_{T}^{ij}\beta_{T}^{pq}-\tilde{k}_{A}^{2}R_{T}^{2}(\beta_{T}^{ij}\beta_{T}^{pq}+2\alpha_{T}^{ij}\beta_{T}^{pq})+\tilde{k}_{A}^{4}R_{T}^{4}\alpha_{T}^{ij}\alpha_{T}^{pq}]\cos{(2\tilde{k}_{A}R_{T})}+\frac{4\tilde{\mathcal{U}}_{ijpq}}{R_{T}^{5}}\tilde{k}_{A}[\beta_{T}^{ij}\beta_{T}^{pq}\nonumber\\
&-\tilde{k}_{A}^{2}R_{T}^{2}\alpha_{T}^{ij}\beta_{T}^{pq}]\sin{(2\tilde{k}_{A}R_{T})}-\frac{2\tilde{\mathcal{U}}_{ijpq}}{R_{T}^{6}}[\beta_{T}^{ij}\beta_{T}^{pq}-k_{B}^{2}R_{T}^{2}(\beta_{T}^{ij}\beta_{T}^{pq}+2\alpha_{T}^{ij}\beta_{T}^{pq})+k_{B}^{4}R_{T}^{4}\alpha_{T}^{ij}\alpha_{T}^{pq}]\nonumber\\
&\times\cos{(2k_{B}R_{T}+\tilde{\Delta}_{AB}T)}-\frac{4\tilde{\mathcal{U}}_{ijpq}}{R_{T}^{5}}k_{B}[\beta_{T}^{ij}\beta_{T}^{pq}-k_{B}^{2}R_{T}^{2}\alpha_{T}^{ij}\beta_{T}^{pq}]\sin{(2k_{B}R_{T}+\tilde{\Delta}_{AB}T)},
\end{align}
which corresponds to the result obtained in Ref.\cite{PRLvdW} for $\langle W_{A}(T)\rangle^{\mathbf{0}}$ but for the replacement of $k_{A}$ and $\Delta_{AB}$  with their Doppler shifted values,  $k_{A}\rightarrow \tilde{k}_{A}=k_{A}(1-v_{R}/c)$, $\Delta_{AB}=\omega_{A}-\omega_{B}\rightarrow\tilde{\Delta}_{AB}=\omega_{A}-\omega_{B}(1+v_{R}/c)$ and $\mathcal{U}_{ijpq}\rightarrow\tilde{\mathcal{U}}_{ijpq}=\mu^{A}_{i}\mu^{A}_{q}\mu^{B}_{j}\mu^{B}_{p}/[(4\pi\epsilon_{0})^{2}\hbar\tilde{\Delta}_{AB}]$. Finally, the lag term reads
\begin{align}\label{Lag}
\langle W_{A}(T)\rangle^{\mathbf{v}}_{Lag}&=(v_{R}/R_{T}\Delta_{AB})2\mathcal{U}_{ijpq}\bigl\{[2c^{-1}\Delta_{AB}R_{T}\cos{(2k_{A}R_{T})}-\sin{(2k_{A}R_{T})}][3\beta_{T}^{ij}\beta_{T}^{pq}/R_{T}^{6}+2\frac{k^{2}_{A}}{R_{T}^{4}}(\beta_{T}^{ij}\beta_{T}^{pq}+2\alpha_{T}^{ij}\beta_{T}^{pq})\nonumber\\&+k^{4}_{A}\alpha_{T}^{ij}\alpha_{T}^{pq}/R_{T}^{2}]-[2c^{-1}\Delta_{AB}R_{T}\sin{(2k_{A}R_{T})}+\cos{(2k_{A}R_{T})}](5k_{A}\beta_{T}^{ij}\beta_{T}^{pq}/R_{T}^{5}-3k_{A}^{3}\alpha_{T}^{ij}\beta_{T}^{pq}/R_{T}^{3})\bigr\},
\end{align}
\end{widetext}
where oscillating transient terms of frequency much higher than $\Delta_{AB}$ have been omitted for simplicity. 

Next, we proceed to compute the velocity-dependent vdW force on atom $A$ at order $v/c$, $\mathbf{F}^{vdW}_{A\mathcal{O}(v)}$, which is the gradient of the difference between the electric interaction at velocity $\mathbf{v}\neq\mathbf{0}$ and at zero velocity,
\begin{align}
\langle\mathbf{F}^{vdW}_{A\mathcal{O}(v)}\rangle&=-\frac{1}{2}\mathbf{\nabla}_{\mathbf{R}_{T}}[\langle W^{E}_{A}(T)\rangle^{\mathbf{v}}_{Dop}-\langle W^{E}_{A}(T)\rangle^{\mathbf{0}}\nonumber\\&+\langle W^{E}_{A}(T)\rangle^{\mathbf{v}}_{\Theta}+\langle W^{E}_{A}(T)\rangle^{\mathbf{v}}_{Lag}].
\end{align}
Averaging in time Eq.(\ref{Dop}) for $T\gg|\Delta_{AB}^{-1}|$, averaging over dipole orientations and restricting ourselves to the near-field and far-field regimes, for which $k_{A}R_{T}\ll1$ and $k_{A}R_{T}\gg1$ respectively, we are left with
\begin{align}
\langle\mathbf{F}^{vdW}_{A\mathcal{O}(v)}\rangle&\simeq-20\frac{\mathcal{U}}{R_{T}^{7}}\frac{(k_{A}+k_{B})}{\Delta_{AB}}\mathbf{v}_{R},\:k_{A}R_{T}\ll1,\nonumber\\
\langle\mathbf{F}^{vdW}_{A\mathcal{O}(v)}\rangle&\simeq4\frac{\mathcal{U}k_{A}^{5}}{9R_{T}^{2}}\frac{k_{B}}{\Delta_{AB}}\mathbf{v}_{R}
[\sin{(2k_{A}R_{T})}\label{ForceB}\\&-2c^{-1}\Delta_{AB}R_{T}\cos{(2k_{A}R_{T})}],\:\:k_{A}R_{T}\gg1,\nonumber
\end{align}
where $\mathcal{U}=\mu^{2}_{A}\mu_{B}^{2}/[(4\pi\epsilon_{0})^{2}\hbar\Delta_{AB}]$. Interestingly, while in the near-field both Doppler and lag terms contribute at the same order to $\langle\mathbf{F}^{vdW}_{A\mathcal{O}(v)}\rangle$, Doppler terms dominate in the far-field. Terms coming from $\langle W^{E}_{A}(T)\rangle^{\mathbf{v}}_{\Theta}$ are negligible in both cases. In addition, while $\langle\mathbf{F}^{vdW}_{A\mathcal{O}(v)}\rangle$ is antiparallel to $\mathbf{v}_{R}$ in the near-field so that it can be referred to as a friction force, it oscillates in space in the far field. Physically, $\langle\mathbf{F}^{vdW}_{A\mathcal{O}(v)}\rangle$ corresponds to the velocity-dependent recoil experienced by atom $A$ during the processes of absorption and emission of the photons that it exchanges with atom $B$, which takes place at a rate $|\Delta_{AB}|$. $\langle\mathbf{F}^{vdW}_{A\mathcal{O}(v)}\rangle$ is indeed proportional to the adimensional factor $v_{R}\omega_{A,B}/c\Delta_{AB}$, and hence to the component of the decay rate of atom $A$ which depends on the presence of atom $B$, $\mathbf{F}^{vdW}_{A}=-\sum_{\mathbf{\epsilon}}\int d^{3}k\:\hbar\mathbf{k}\frac{d}{2dT}(|\langle A_{-}B_{-},\gamma_{\mathbf{k},\mathbf{\epsilon}}|\mathbb{U}(T)|\Psi(0)\rangle|^{2})$ \footnote{Denoting by $_{\Psi(0)}P_{A_{-}B_{-}\gamma}(T)$ the probability of emission of a single photon at a time $T$ after the excitation of atom $A$,  $_{\Psi(0)}P_{A_{-}B_{-}\gamma}(T)=\int d^{3}k\sum_{\mathbf{\epsilon}}\langle\Psi(0)|\mathbb{U}^{\dagger}(T)|A_{-}B_{-},\gamma_{\mathbf{k},\mathbf{\epsilon}}\rangle\langle A_{-}B_{-},\gamma_{\mathbf{k},\mathbf{\epsilon}}|\mathbb{U}(T)|\Psi(0)\rangle$, the following relation holds at our approximation order,\\$\langle W^{E}_{A}(T)\rangle=-i\hbar\int d^{3}k\sum_{\mathbf{\epsilon}}\frac{d}{dT}\Bigl(\langle\Psi(0)|\mathbb{U}^{\dagger}(T)|A_{-}B_{-},\gamma_{\mathbf{k},\mathbf{\epsilon}}\rangle\Bigr)$ $\times\langle A_{-}B_{-},\gamma_{\mathbf{k},\mathbf{\epsilon}}|\mathbb{U}(T)|\Psi(0)\rangle+\:c.c.$,   with  $|A_{-}B_{-},\gamma_{\mathbf{k},\mathbf{\epsilon}}\rangle$ being a single photon state of momentum $\hbar\mathbf{k}$ and polarization $\epsilon$. From here we have that $\mathbf{F}^{vdW}_{A}=-\sum_{\mathbf{\epsilon}}\int d^{3}k\:\hbar\mathbf{k}\frac{d}{2dT}|\langle A_{-}B_{-},\gamma_{\mathbf{k},\mathbf{\epsilon}}|\mathbb{U}(T)|\Psi(0)\rangle|^{2}.\:\:\qquad$}. The expression within parenthesis is the probability of emission of a single photon of momentum $\hbar\mathbf{k}$ and polarization $\mathbf{\epsilon}$ at a time $T$ after the excitation of atom $A$.

Finally, for the sake of completeness we compute the velocity-dependent vdW force on the excited atom $A$ as it flies with velocity $\mathbf{v}$ parallel to a thin plate ($\Pi$) parallel to the $xy$ plane and made of a random distribution of ground state atoms of the kind $B$. The numerical atomic surface density is denoted by $\sigma$, and the distance of atom $A$ to the plate by $d$ --see Fig.\ref{figure1}(b). For simplicity, we adopt the pairwise summation approximation, according to which the total force over atom $A$ is the sum of all the pair forces between atom $A$ and each atom $B$. Equipped with our computation of $\langle\mathbf{F}^{vdW}_{A\mathcal{O}(v)}\rangle$, the problem becomes simply geometrical, and its result is 
\begin{align}
\langle\mathbf{F}_{A\mathcal{O}(v)}^{vdW}\rangle_{\Pi}&\simeq\sigma\int_{-\infty}^{\infty}\textrm{d}x\int_{-\infty}^{\infty}\textrm{d}y\langle\mathbf{F}_{A\mathcal{O}(v)}^{vdW}(x,y,d)\rangle\nonumber\\
&=-8\pi\sigma\frac{\mathcal{U}}{21d^{5}}\frac{(k_{A}+k_{B})}{\Delta_{AB}}\mathbf{v},\:k_{A}d\ll1,\nonumber\\
\langle\mathbf{F}_{A\mathcal{O}(v)}^{vdW}\rangle_{\Pi}&\simeq-2\pi\sigma\frac{\mathcal{U}k_{A}^{3}}{9d^{2}}\frac{k_{B}}{\Delta_{AB}}\mathbf{v}[\sin{(2k_{A}d)}\nonumber\\&-2c^{-1}\Delta_{AB}d\cos{(2k_{A}d)}],\:\: k_{A}d\gg1.\label{ForcePi}
\end{align}
In contrast to the result of Ref.\cite{BuhmannScheel}, $\langle\mathbf{F}_{A\mathcal{O}(v)}^{vdW}\rangle_{\Pi}$ is always a friction force for $k_{A}d\ll1$, while the direction of the force of Eq.(79) in Ref.\cite{BuhmannScheel} depends on the sign of $\Delta_{AB}$. However, this essential discrepancy is due to the particular form of the effective permittivity adopted by the authors to characterize the plate there --i.e., a Drude-Lorentz form, and not to an essential difference between their approach and ours. As a matter of fact, inserting the  polarisability of atom B in the scattering Green's function of Eq.(65) in Ref.\cite{BuhmannScheel}  and adopting the quasi-resonant approximation, a term is obtained of the same order as that in our Eq.(\ref{ForceB}) in the non-retarded regime, with different numerical prefactors though.

\section{R\"ontgen force}\label{Rontgen}

In this section we consider the force induced by the coupling to the R\"ontgen momentum and the force due to the time variation of the R\"ontgen momentum itself. They correspond in Eq.(\ref{Force}), respectively, to conservative and non-conservative forces.

As for the conservative force induced by the inclusion of the interaction $W_{A}^{R}$ in Eq.(\ref{Force}), it does not only yield an additional potential energy, $\langle W_{A}^{R}\rangle$, but also contributes  to $\langle W_{A}^{E}\rangle$ through the insertion of $W_{A}^{R}$ in the time-evolution operators. Since $W_{A}^{R}$ is already linear in $\langle\mathbf{Q}_{A}\rangle$, it provides terms of orders $v/c$, $(v/c)^{3}$ and higher to the conservative force  without the need of considering the quantum nature of the CM d.o.f. within the EM operators. Thus, an analogous formula to that for $\langle W_{A}^{E}(T)\rangle$ in Eq.(\ref{laeq}) but for the replacement of one of the operators $W_{A}^{E}$ by $W_{A}^{R}$  yields the contribution of $W_{A}^{R}$ to the interaction energy. Upon applying the operator $-\frac{1}{2}\mathbf{\nabla}_{\mathbf{R}_{T}}$ we find, at order $\mathcal{O}(v/c)$,  
\begin{align}
&\langle\mathbf{F}^{R-c}_{A}\rangle\simeq\frac{-2\alpha_{f}c^{2}}{\pi\epsilon_{0}e^{2}\Delta_{AB}}\mu^{A}_{i}\mu^{B}_{j}\mu^{B}_{p}\mu^{A}_{q}\epsilon^{i}_{\:lm}v^{l}\mathbf{\nabla}_{\mathbf{R}_{T}}\int_{-\infty}^{+\infty}\textrm{d}k\:k^{2}\nonumber\\
&\times\frac{\textrm{Im}\{\mathcal{G}^{mj}(kR_{T})\}}{k-k_{A}-i\eta}\int_{-\infty}^{+\infty}\textrm{d}k'k'^{2}\frac{\textrm{Im}\{G^{pq}(k'R_{T})\}}{k'-k_{A}-i\eta}+\textrm{c.c.},\nonumber
\end{align}
where the superscript $R-c$ in $\langle\mathbf{F}_{A}^{R-c}\rangle$ denotes 'R\"ontgen-conservative', $\epsilon_{ilm}$ is the three-dimensional Levi-Civita tensor and the causal-adiabatic approximation has been considered with $\eta\rightarrow0^{+}$. As explained in Ref.\cite{PRLvdW}, the latter amounts to considering averages in time over a time-interval much greater than $|\Delta_{AB}^{-1}|$.  Mathematically, this is achieved by replacing each interaction Hamiltonian $W$ by $We^{\eta t}$ within the time integrals, with $\eta\rightarrow0^{+}$, and by extending the lower limits of integration to $-\infty$. In the above equation $\mathcal{G}(kR)$ is the dyadic Green's function of the magnetic field induced at a distance $R$ by an electric dipole of frequency $ck$ in free space. Their components are
\begin{equation}
\mathcal{G}^{mj}(kR)=\frac{e^{ikR}}{4\pi cR}(1+i/kR)\epsilon^{msj}R_{s}/R.
\end{equation}
After performing the frequency integrals in the complex plane we end up with
\begin{align}
\langle\mathbf{F}^{R-c}_{A}\rangle&=\frac{-4\pi\alpha_{f}ck_{A}^{4}}{\epsilon_{0}e^{2}\Delta_{AB}}\mu^{A}_{i}\mu^{B}_{j}\mu^{B}_{p}\mu^{A}_{q}\epsilon^{i}_{lm}v^{l}\nonumber\\
&\times\mathbf{\nabla}_{\mathbf{R}_{T}}\textrm{Re}\{\mathcal{G}^{mj}(k_{A}R_{T})G^{pq}(k_{A}R_{T})\}.\nonumber
\end{align}
 Straightforward evaluation of the Green's functions at $k_{A}R_{T}$ reveals that, in the near field,  $\langle\mathbf{F}^{R-c}_{A}\rangle/\langle\mathbf{F}^{vdW}_{A\mathcal{O}(v)}\rangle\sim|\Delta_{AB}|k_{A}R_{T}^{2}/c\ll1$, whereas in the far field the ratio goes like $\sim|\Delta_{AB}|/ck_{A}\ll1$. From this we conclude that the conservative R\"ontgen force is always negligible.

As for the non-conservative ($R-nc$) force, 
\begin{equation}
\langle\mathbf{F}^{R-nc}_{A}\rangle=\partial_{T}\langle\Psi(0)|\mathbb{U}^{\dagger}(T)\mathbf{d}_{A}\times\mathbf{B}(\mathbf{R}_{A})\mathbb{U}^{\dagger}(T)|\Psi(0)\rangle\label{RontT}
\end{equation}
corresponds to the time-derivative of the R\"ontgen momentum. It vanishes at zero velocity and an analogous calculation to the one for  $\langle\mathbf{F}^{vdW}_{A\mathcal{O}(v)}\rangle$ must be performed for $\mathbf{v}\neq\mathbf{0}$, i.e., the CM d.o.f. must be promoted to quantum variables. The only differences w.r.t. the calculation of $\langle\mathbf{F}^{vdW}_{A\mathcal{O}(v)}\rangle$ are the replacements of an operator $W_{A}^{E}$  with $-\mathbf{d}_{A}\times\mathbf{B}(\mathbf{R}_{A})$ and of $-\frac{1}{2}\mathbf{\nabla}_{\mathbf{R}_{T}}$ with $\partial_{T}$. This procedure leads to a formula similar to that in Eq.(\ref{Wv}),
\begin{align}
\langle F^{R-nc}_{A}\rangle_{i}&=\frac{2\alpha_{f}c^{3}}{\pi\epsilon_{0}e^{2}}\epsilon_{i}^{\:\:rp}\mu^{A}_{r}\mu^{B}_{q}\mu^{B}_{m}\mu^{A}_{n}\int_{-\infty}^{+\infty}\int\textrm{d}k'\textrm{d}k\:k'^{2}k^{2}\nonumber\\
&\times\textrm{Re}\:\partial_{T}\int_{-\infty}^{T}\textrm{d}t\int_{-\infty}^{t}\textrm{d}t'\int_{-\infty}^{t'}\textrm{d}t''e^{\eta(t+t'+t'')}\Bigl\{\Bigl[\nonumber\\
&\times\bigl[\textrm{Im}\{G^{mn}(k'R_{T})\}\textrm{Im}\{\mathcal{G}^{pq}[k|\mathbf{R}_{T}-\mathbf{v}(T-t'')|]\}\nonumber\\&+\textrm{Im}\{\mathcal{G}^{pq}(k'R_{T})\}\textrm{Im}\{G^{mn}[k|\mathbf{R}_{T}-\mathbf{v}(T-t'')|]\}\bigr]\nonumber\\
&\times i\:e^{i\omega_{A}T}e^{-i(T-t)\omega'}e^{-i(t-t')\omega_{B}}e^{-i(t'-t'')\omega}e^{-it''\omega_{A}}\Bigr]\nonumber\\&+[\omega\leftrightarrow\omega']\Bigr\},\label{FWRon}
\end{align}
where the causal-adiabatic approximation has been applied with $\eta\rightarrow0^{+}$. Same as for Eq.(\ref{GDoppLag}), the $\mathbf{v}$-dependent Green's functions within the integrand can be split into Doppler terms and lag terms linear in $\mathbf{v}$. The Doppler terms yield null contribution to the R\"ontgen force. The lag terms of the dyadic Green's functions are, in addition to the second term of Eq.(\ref{GDoppLag}) from $\mathbb{G}[k|\mathbf{R}_{T}-\mathbf{v}(T-t'')|]$, 
\begin{align}
\delta G^{Lag,R}_{nm}(kR_{T})&=v_{R}(T-t'')\frac{ke^{ikR_{T}}}{4\pi R_{T}}\Bigl[\frac{\alpha_{T,nm}}{kR_{T}}+\frac{2i\beta_{T,nm}}{(kR_{T})^{2}}\nonumber\\
&-\frac{3\beta_{T,nm}}{(kR_{T})^{3}}\Bigr],\nonumber
\end{align}
a term coming from  $\mathcal{G}[k|\mathbf{R}_{T}-\mathbf{v}(T-t'')|]$ which is also proportional to $v_{R}(T-t'')/R_{T}$,
\begin{equation}
\delta\mathcal{G}^{Lag,R}_{pq}(kR_{T})=v_{R}(T-t'')\frac{ke^{ikR_{T}}}{4\pi cR_{T}}\left(\frac{1}{kR_{T}}+\frac{2i}{k^{2}R^{2}_{T}}\right)\epsilon_{\:\:s}^{p\:\:q}\frac{R^{s}_{T}}{R_{T}},\nonumber
\end{equation}
and two more terms proportional to $v_{\perp}(T-t'')/R_{T}$ coming from the expansion of the unit radial vector within the tensors $\alpha,\beta$ in $\mathbb{G}[k|\mathbf{R}_{T}-\mathbf{v}(T-t'')|]$ and within $\mathcal{G}[k|\mathbf{R}_{T}-\mathbf{v}(T-t'')|]$,
\begin{align}
&\delta G^{Lag\perp}_{nm}(kR_{T})=(T-t'')\frac{k\:e^{ikR_{T}}}{4\pi R_{T}}[1/kR_{T}+3i/(kR_{T})^{2}\nonumber\\
&-3/(kR_{T})^{3}](\delta^{i}_{n}R_{T,i}R^{-1}_{T}\delta^{j}_{m}v_{\perp,j}+\delta^{i}_{n}v_{\perp,i}\delta^{j}_{m}R_{T,j}R^{-1}_{T}),\nonumber\\
&\delta\mathcal{G}^{Lag\perp}_{pq}(kR_{T})=-(T-t'')\frac{ke^{ikR_{T}}}{4\pi cR_{T}}(1/kR_{T}+i/k^{2}R^{2}_{T})\epsilon_{\:\:s}^{p\:\: q}v^{s}_{\perp}\nonumber
\end{align}
respectively, with $\mathbf{v}_{\perp}$ being the component of $\mathbf{v}$ perpendicular to $\mathbf{R}_{T}$. The reason for the neglect of $\delta\mathbb{G}^{Lag\perp}$ in the calculation of $\langle\mathbf{F}^{vdW}_{A\mathcal{O}(v)}\rangle$ is that its contribution does not survive orientational average there, whereas it does so in $\langle\mathbf{F}^{R-nc}_{A}\rangle$. Finally, performing the time and frequency integrals of Eq.(\ref{FWRon}) we are left with
\begin{align}
&\langle F^{R-nc}_{A}\rangle_{i}=\frac{-2k_{A}^{5}\alpha_{f}c^{2}\pi}{\Delta_{AB}^{2}\epsilon_{0}e^{2}}\epsilon_{i}^{\:\:rp}\frac{\mu^{A}_{r}\mu^{B,q}\mu^{B}_{m}\mu^{A}_{n}}{(T-t'')}\Bigl\{6G^{mn}(k_{A}R_{T})\nonumber\\
&\times[\delta\mathcal{G}^{Lag,R}_{pq}(k_{A}R_{T})+\delta\mathcal{G}^{Lag\perp}_{pq}(k_{A}R_{T})]\nonumber\\
&+k_{A}\Bigl[G^{mn}(kR_{T})[\delta\mathcal{G}^{Lag,R}_{pq}(kR_{T})+\delta\mathcal{G}^{Lag\perp}_{pq}(kR_{T})]\Bigr]'_{k=k_{A}}\Bigr\}\nonumber\\
&+\{G\rightarrow\delta G,\:\delta\mathcal{G}\rightarrow\mathcal{G}\},
\end{align}
where the prime on the big square brackets denotes the derivative w.r.t. $k$ of the functions therein. Note that, in contrast to its cancellation found in Ref.\cite{BuhmannScheel} for an atom in an incoherent superposition of states, vanishing of $\langle\mathbf{F}^{R-nc}_{A}\rangle$ does not hold for an atom in a metastable excited state. On physical grounds, a non-vanishing value of the R\"ontgen momentum must be expected for a polarizable atom $A$ in motion and subject to the electric and magnetic fields induced by a second atom $B$ at rest \cite{JPMC}. Less obvious is the fact that the R\"ontgen momentum varies in time.

Straightforward evaluation of the Green's functions in the equation above reveals that, in the near field,  $\langle\mathbf{F}^{R-nc}_{A}\rangle/\langle\mathbf{F}^{vdW}_{A\mathcal{O}(v)}\rangle\sim k_{A}^{2}R_{T}^{2}\ll1$ and hence it is negligible. On the contrary, in the retarded regime, after performing the orientational average over the random orientation of the electric dipoles, we find
\begin{equation}
\langle\mathbf{F}^{R-nc}_{A}\rangle\simeq\frac{2k_{A}^{6}\mathcal{U}}{9R_{T}^{2}\Delta_{AB}}\sin{(2k_{A}R_{T})}[2\mathbf{v}_{R}-\mathbf{v}_{\perp}],\: k_{A}R_{T}\gg1,\label{FRont}\\
\end{equation}
whose radial component equals $\langle\mathbf{F}^{vdW}_{A\mathcal{O}(v)}\rangle$.

Finally, as we did for the velocity-dependent vdW force we compute here the R\"ontgen force on the excited atom $A$ as it flies with velocity $\mathbf{v}$ parallel to a thin plate ($\Pi$) of ground state atoms of the kind $B$, at a distance $d$. Assuming that $k_{A}d\gg1$,  we obtain
\begin{align}
\langle\mathbf{F}^{R-nc}_{A}\rangle_{\Pi}&\simeq\pi\sigma\frac{-2\mathcal{U}k_{A}^{5}}{9d\Delta_{AB}}\mathbf{v}[\cos{(2k_{A}d)}+2\sin{(2k_{A}d)}/k_{A}d].\label{ForcePiR}
\end{align}
The dominant term in this equation comes from the integration over the plane of the component of $\langle\mathbf{F}^{R-nc}_{A}\rangle$ which is parallel to $\mathbf{v}_{\perp}$. Interestingly, this term is an order $k_{A}d$ or $\omega_{A}/\Delta_{AB}$ greater than the velocity-dependent vdW force in the retarded regime.

\section{Conclusions}\label{Conclusion}
We have performed the time-dependent calculation of the velocity-dependent  forces acting on an excited atomic dipole which moves at constant velocity with respect to ground state atoms of a different kind,  in the quasiresonant approximation at zero temperature. To this aim, we have used time-dependent quantum perturbation theory. Both the interaction with a single ground state atom and with a thin plate made of a random distribution of independent atoms have been evaluated. We find that, at leading order, these interactions are linear in the velocity and no relativistic corrections enter at this order.

In either case, the total force consists of a conservative van der Waals  component and a non-conservative R\"ontgen component. For their computation, the position and kinetic momentum vectors of the CM of the atoms have been promoted to quantum variables. We have shown that relativistic corrections to the van der Waals force enter at order $(v/c)^{5}$. 

In the non-retarded regime the van der Waals force dominates, being always antiparallel to the velocity --hence can be referred to as quantum friction.
On physical grounds, this force corresponds to the velocity-dependent recoil experienced by the excited atom during the processes of absorption and emission of the photons that it exchanges with the ground state atoms, which takes place at a rate $|\Delta_{AB}|$. This force is indeed proportional to $v\omega_{A,B}/c\Delta_{AB}$ and hence to the emission rate of the excited atom which is induced by the ground state atoms. 

On the contrary, in the retarded regime the velocity-dependent forces oscillate in space, the van der Waals and the R\"ontgen forces are of the same order in the atom-atom interaction, and the R\"ontgen component dominates in the atom-surface interaction. The latter corresponds to the time-variation of the R\"ontgen momentum, which is also mediated by the periodic exchange of quasi-resonant photons.

\acknowledgments
We thank Stefan Scheel and Diego Dalvit for fruitful discussions. We are also grateful to the referee for having encouraged us to investigate the effect of the R\"ontgen current. Financial support from ANR-10-IDEX-0001-02-PSL and ANR-13-BS04--0003-02 is gratefully acknowledged.

\end{document}